\documentclass[sn-apa]{sn-jnl}

\usepackage{graphicx}%
\usepackage{multirow}%
\usepackage{amsmath,amssymb,amsfonts}%
\usepackage{amsthm}%
\usepackage{mathrsfs}%
\usepackage[title]{appendix}%
\usepackage{xcolor}%
\usepackage{textcomp}%
\usepackage{manyfoot}%
\usepackage{booktabs}%
\usepackage{algorithm}%
\usepackage{algorithmicx}%
\usepackage{algpseudocode}%
\usepackage{listings}%

\theoremstyle{thmstyleone}%
\theoremstyle{thmstyletwo}%

\theoremstyle{thmstylethree}%

\begin{document}

\title[Article Title]{The impact of gamification on learning outcomes: experiences from a Biomedical Engineering course}


\author*[1,2]{\fnm{Gonzalo R.} \sur{R\'ios-Mu\~{n}oz}}\email{grios@ing.uc3m.es} 

\author[1,2]{\fnm{Caterina} \sur{Fuster-Barcel\'o}}\email{cafuster@pa.uc3m.es} 

\author[1,2]{\fnm{Arrate} \sur{Mu\~{n}oz-Barrutia} }\email{mamunoz@ing.uc3m.es} 


\affil*[1]{\orgdiv{Bioengineering Department}, \orgname{Universidad Carlos III de Madrid}, \orgaddress{\street{Avenida de la Universidad 30}, \city{Legan\'es}, \postcode{28911}, \state{Madrid}, \country{Spain}}}

\affil[2]{\orgname{Instituto de Investigaci\'on Sanitaria Gregorio Mara\~{n}\'on}, \orgaddress{\street{Calle Dr. Esquerdo 46}, \city{Madrid}, \postcode{28007}, \state{Madrid}, \country{Spain}}}











\abstract{
This study examines the integration of digital tools in project-based learning within a Biomedical Engineering course to enhance collaboration, transparency, and assessment fairness. Building on prior pilot experiences, we implemented a structured learning environment that combined experiment tracking, real-time collaboration, and peer-assessment practices. The intervention was deployed across two consecutive academic years, involving master’s-level students in Biomedical Image Processing. Data were collected through project outcomes, peer-assessment rubrics, and student surveys. Results show that the integration of digital platforms supported accountability, improved the quality of collaborative work, and fostered greater equity in the evaluation process. Students highlighted increased engagement, enhanced teamwork, and clearer criteria for performance assessment. Faculty reported more efficient monitoring of progress and improved feedback practices. Despite challenges such as technical adoption and the need for instructor guidance, the study demonstrates the potential of structured tool integration to support active and transparent learning environments. Findings contribute to the broader discourse on digital pedagogy, offering a replicable model for higher education contexts in science and technology.
}




\maketitle

\section{Introduction}\label{sec1}



The digital age has transformed how information is consumed and processed, especially among younger generations, reshaping contemporary education. The digital nativity of these cohorts presents both a challenge and an opportunity for educational methodologies to evolve in tandem with technological advancements and changing learner behaviors \cite{D_Bekebrede2011}. The traditional didactic approaches, characterized by their unidirectional flow of information and reliance on rote learning, are increasingly seen as misaligned with the dynamic, interactive, and information-rich environments that digital natives are accustomed to \cite{D_Blocksidge2023}. 

The imperative for educational reform is not just about adopting new technologies; it is about reimagining the pedagogical framework to be more engaging, adaptive, and reflective of real-world complexities. The transition from conventional teaching methods to innovative pedagogical strategies is pivotal in enhancing student engagement, facilitating deeper comprehension, and promoting the application of knowledge \cite{D_Kiser2015}. This evolution is crucial not only for improving academic outcomes but also for ensuring higher retention rates and fostering a lifelong learning mindset among students.

Recent pedagogical innovations have explored the integration of online learning platforms, such as Engageli, Google Classroom, and Zoom, to enhance the flexibility and accessibility of education, allowing learners to engage with content at their own pace and from any location \cite{D_Barak2016, D_Sofi-Karim2023, D_Dash2022}. Moreover, the exploration of mobile learning applications and augmented reality technologies has opened new avenues for creating interactive and immersive learning experiences that can captivate student interest and facilitate a deeper understanding of complex subjects \cite{D_Oliveira2021, D_Domingo2016, D_Atwood-Blaine2017, D_Quintero2019, D_Wei2021, D_Bi2019}.

As we venture into these new educational territories, it is crucial to not only update the tools and technologies we use but also to reconsider our engagement strategies with students. Adhering to outdated pedagogical models risks diminishing the efficacy of the learning experience \cite{D_Ong2023, D_Sun2022}. In this context, gamification emerges as a promising approach, providing a framework for making learning more engaging and interactive while also preparing students to develop entrepreneurial skills and apply theoretical knowledge in practical settings \cite{D_Sanchez2020, D_Deterding2011, D_Mei2022}. However, the challenge remains to integrate these innovative strategies effectively within technical and higher education curricula, often criticized for their heavy focus on theoretical content at the expense of practical skills development and professional identity formation \cite{D_Hua2022}. 
 
This paper aims to present a novel pedagogical framework that incorporates motivational and entrepreneurial methodologies within a nanotechnology course, demonstrating its applicability and versatility across various Science, Technology, Engineering and Mathematics (STEM) disciplines. We begin by reviewing existing teaching methodologies and revisit the "Theory of Gamified Learning" proposed by \cite{D_Landers_2014} that provides a structured approach to integrating gamification in educational settings. It offers a detailed framework that explains how different game elements can influence learner motivation, engagement, cognitive development, and social interaction. By applying this theory, we can design and analyze more engaging and effective learning experiences that leverage the motivational and interactive potential of gamification and serious games. Then, we introduce our adaptable framework through a detailed discussion of the course structure. We propose an evaluation system grounded in gamification and entrepreneurial activities and outline methods to assess the impact of our approach on student satisfaction and academic performance, and we relate it to the theoretical structured approach proposed by \cite{D_Landers_2014}. Our findings and conclusions will shed light on the effectiveness of this teaching framework in enhancing STEM education. 

By revisiting and revitalizing the pedagogical approaches within STEM education, we can better align with the preferences and expectations of modern learners, fostering an educational environment that is not only informative but also engaging, practical, and conducive to lifelong learning. 

\section{Gamification and Engaging in Higher-Level Education}

Traditional pedagogical models in higher education, particularly within the STEM disciplines, have predominantly relied on lecture-based teaching methods characterized by a unidirectional flow of information from educator to student. This conventional approach, while structured, often results in a passive learning experience, limiting the ability of the students to engage critically with the content and to apply theoretical concepts to real-world problems \cite{D_Guerra2017}. The limitations of such methods have become increasingly apparent, prompting a reevaluation of educational strategies to accommodate better the needs of a digitally savvy and diverse student body. 

Empirical research within the field of STEM education has consistently highlighted the benefits of active learning strategies over traditional lecture-based approaches. Studies have shown that active learning not only enhances student engagement and comprehension but also significantly improves academic performance across various STEM fields \cite{D_Freeman2014}. The effectiveness of these strategies challenges the continued reliance on traditional lecturing as a baseline pedagogical method and advocates for a paradigm shift towards more interactive student-centered learning models \cite{D_Hernandez2019}.

In response to this evolving educational landscape, there has been a marked increase in the adoption of gamification as a pedagogical tool to promote active learning within engineering and other STEM disciplines \cite{D_Hernandez2019}. Gamification, the application of game-design elements in non-game contexts, has been shown to foster an engaging and dynamic learning environment, encouraging students to actively participate and apply their knowledge in practical settings \cite{D_Gasca2021, C_Su2019, D_Popovic2018, D_Alhammad2018, D_Souza2018, D_Pedreira2015}. Although nowadays gamification likely suffers from the novelty effect, it also benefits from the familiarization effect \cite{D_Rodrigues_2022}. The success of gamification has been documented across a range of technical fields, from Thermal and Construction Engineering to Software and Chemical Engineering, demonstrating its versatility and efficacy in enhancing educational outcomes \cite{D_SuarezLopez2023, D_Ilbeigi2022, D_Ngandu2023, D_Paniagua2019}.

Research into electronic gamification (e-gamification) further supports the potential of digital tools to increase student engagement and participation in STEM education \cite{D_Cuevas2019, D_Nowostawski2018}. E-gamification platforms offer immersive experiences that can captivate the attention of the students, foster collaborative learning, and provide instructors with valuable insights into student progress \cite{D_Murillo2021}, with assessment approaches mimicking traditional video games such as Super Mario \cite{D_Cuevas2019}. 

In exploring alternative e-gamification tools, researchers have delved into the utilization of mobile applications, quizzes, and augmented reality within educational settings. It was found that the selections of mobile apps by educators are closely linked to their perceptions of the impact that these technologies have on student learning outcomes \cite{D_Domingo2016}. Conversely, a separate study revealed that students perceive the use of mobile applications during online classes as potentially distracting and often misjudge the amount of time they allocate to these applications \cite{D_Oliveira2021}. Regarding quizzes and surveys, the findings indicate that these tools significantly enhance the cognitive engagement of the students during the learning process \cite{D_Mayer2009} and serve as indicators of heightened learning engagement \cite{D_Ong2023}. Moreover, current technological advancements are probing the potential of augmented reality in a pedagogical context \cite{D_Quintero2019}. Notably, \cite{D_Wei2021} highlight the innovative application of augmented reality in education, underscoring its capacity to enhance both learning and knowledge acquisition proudly. Augmented reality is identified as a groundbreaking tool within the educational domain, markedly enhancing the critical thinking of the students, their knowledge assimilation, their engagement, and their overall academic achievement \cite{D_Bi2019}.

Engagement and motivation are not merely pivotal in gamification strategies; they are foundational elements of successful gamification deployments that potentially influence the academic achievements of students. Literature indicates that the effective incorporation of gaming elements into assessment processes has led to notable improvements in scores and completion rates \cite{D_Cuevas2019, D_Paniagua2019}, accompanied by heightened student satisfaction \cite{D_SuarezLopez2023, D_Murillo2021, D_Abuhassna2020}. Furthermore, the interaction between teachers and students emerges as a significant determinant of learning efficacy, enhancing student learning through a supportive psychological environment, heightened engagement, and increased motivation. In this realm, educators have observed positive outcomes from leveraging gamification to boost student motivation and engagement, resulting in amplified participation in academic activities \cite{D_Gamarra2022} and enhanced teacher-student interactions, even within the context of online education \cite{D_Sun2022}\cite{D_Ong2023}.

In conclusion, the shift towards gamification and other interactive learning strategies represents a significant advancement in STEM education, aligning teaching methodologies with the preferences and expectations of modern learners. By embracing these innovative approaches, educators can create more engaging, effective, and meaningful learning experiences that prepare students not only for academic success but also for the challenges of the professional world. However, a common and robust theoretical methodology should be implemented to better understand and evaluate the impact of gamification in class \cite{D_Landers_2014, D_Sanchez2020}.

\subsection{Theory of Gamified Learning}

The Theory of Gamified Learning, developed by \cite{D_Landers_2014}, integrates the concepts of serious games and gamification to enhance educational outcomes. In this work, gamification is defined as the application of game attributes in non-game contexts to influence learning-related behaviors or attitudes. This theory proposes that gamification can affect learning through two primary processes: mediation and moderation.

\textbf{Mediation Process:}
In the mediation process, gamified elements are designed to directly influence learner behaviors and attitudes, which in turn, enhance learning outcomes. For instance, the inclusion of points, badges, and leaderboards can increase student engagement and time-on-task, thereby improving knowledge retention and performance. The mediator here is the targeted behavior or attitude, which serves as the conduit through which gamification affects learning outcomes.

\textbf{Moderation Process:}
The moderation process, on the other hand, suggests that gamified elements strengthen the relationship between the quality of instructional content and learning outcomes. In this scenario, gamification enhances the effectiveness of already sound instructional design by making learning activities more engaging. For example, incorporating narrative elements or game-like challenges can increase student motivation, thereby amplifying the impact of the instructional content.

Additionally, the framework provided by \cite{D_Bedwell2012}, which categorizes game attributes into nine categories (e.g., action language, assessment, conflict/challenge, control, environment, game fiction, human interaction, immersion, and rules/goals), is utilized to identify and apply specific game elements in educational settings. Each attribute can be isolated and examined for its individual and combined effects on learning.

The Theory of Gamified Learning is built on five key propositions that outline how instructional content, game characteristics, and learner behaviors interact to influence learning outcomes. These propositions provide a structured approach to understanding the mechanisms through which gamification impacts education.

\begin{itemize}
    \item \textbf{Proposition 1: Instructional content influences learning outcomes and behaviors.}
Improved instructional content can alter learning outcomes and learner behaviors across various contexts.

\item \textbf{Proposition 2: Behaviors/attitudes influence learning.}
Learner attitudes and behaviors significantly impact learning, with constructs like cognitive effort and engagement playing critical roles.

\item \textbf{Proposition 3: Game characteristics influence changes in behavior/attitudes.}
Variation in game characteristics affects learner behaviors and attitudes, which are essential for gamification to achieve its goals.

\item \textbf{Proposition 4: Game elements affect behaviors/attitudes that moderate instructional effectiveness.}
Gamified elements can enhance a learning-related behavior or attitude, which in turn, strengthens the relationship between instructional content and learning outcomes.

\item \textbf{Proposition 5: The relationship between game elements and learning outcomes is mediated by behaviors/attitudes.}
Gamification influences learning outcomes through targeted behaviors or attitudes that serve as mediators in the learning process.
\end{itemize}
    
The theory emphasizes the need for the rigorous, scientific study of gamification in educational contexts. By systematically exploring the impact of individual game elements and their meaningful combinations, researchers can provide clear, actionable insights for instructional designers. The goal is to avoid the pitfalls of poorly designed gamification efforts and to develop evidence-based strategies that effectively enhance learning outcomes.

In summary, the Theory of Gamified Learning provides a comprehensive framework for understanding how gamification can be strategically used to improve educational experiences. It bridges the gap between serious games and gamification, offering a unified approach to leveraging game elements in enhancing learning.

\section{Materials and Methods}

In the specific context of our Bioengineering Bachelor's program, we have implemented a focused teaching methodology within a selected course. This approach integrates dynamic activities, targeted gamification elements, and in-depth project-based learning. Aimed at bridging theoretical knowledge with practical application, this strategy facilitates comprehensive skill development among students. To ensure academic integrity and evaluate the students' overall learning, this method operates in conjunction with traditional exams.

\subsection{Objectives, hypothesis, and Participants}\label{ss:objectives}

\subsubsection{Objectives}
Our study aims to implement a series of cutting-edge strategies to examine student behavior and engagement for subsequent research. The objectives of this research have been established as follows:

\begin{itemize}
    \item \textbf{To assess the impact of two distinct gamification strategies, "Nanogames" and "NanoTechStart," on the academic performance of the students.} "Nanogames," a series of collaborative, interactive games, forms part of the mid-term assessment, while "NanoTechStart" stimulates a start-up fundraising event, challenging students to present a nanotechnology-based solution to a panel of potential investors. 
    
    \item \textbf{To foster student leadership and active participation} in organizing and conducting the "NanoTechStart" event, thereby promoting a deeper engagement with the course material. 
   
    \item \textbf{To leverage gamification as a motivational tool,} enhancing the learning experience and fostering a more dynamic classroom environment. 
    
    \item \textbf{To emphasize the relevance of nanotechnology solutions} to societal and economic challenges, encouraging students to apply their knowledge in addressing real-world problems. 

    \item \textbf{To analyze the differential effects of "Nanogames" and "NanoTechStart" on student cognitive outcomes}, such as problem-solving abilities and critical thinking. This objective involves quantitatively measuring changes in these cognitive skills before and after the intervention, using standardized assessments to identify which aspects of gamification contribute most effectively to cognitive development in a STEM education setting.

\end{itemize}

\subsubsection{Hypotheses}

\begin{figure}[h]
	\centering
		\includegraphics[width=\textwidth]{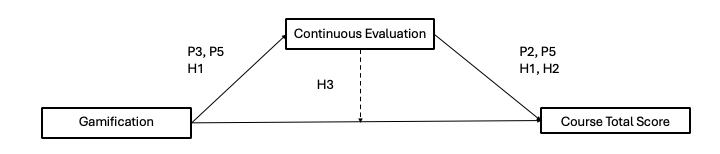}
	  \caption{Theoretical model of the present study based on the gamified theory by \cite{D_Landers_2014}. The model presents a mediation process following the upper path Gamification $\rightarrow$ Continuous Evaluation $\rightarrow$ Course Total Score. The influence of the Continuous evaluation in the Gamification $\rightarrow$ Course Total Score path constitutes a moderating process. Notice that we have not included the instructional content moderator box in the original theoretical model.}
   \label{fig:MMmodel}
   \footnotesize{Notation: H, hypothesis of the present study; P, theory of gamified learning proposition.}
\end{figure}

The effects of gamification on student performance have been rigorously analyzed following the gamification theoretical methodology introduced by \cite{D_Landers_2014}. According to this methodology, three variables interacting in this process have been identified. These are: i) the presence of gamification in the course (independent variable, binary); ii) the course total score (dependent variable, normalized 0-1); iii) and, the performance of the students in the continuous evaluation (mediation-moderation variable, normalized 0-1). From their relationships depicted in Fig.~\ref{fig:MMmodel}, the following hypotheses related to the moderation-mediation model have been identified:

\begin{itemize}
    \item \textbf{Hypothesis 1:} The positive effect of gamification on the total course score will be mediated by continuous evaluation performance.
    \item \textbf{Hypothesis 2:} There will be a strong relationship between the performance of the students in continuous evaluation and the final course score.
    \item \textbf{Hypothesis 3:} Performance in continuous evaluations will moderate the relationship between gamification and total course score.
\end{itemize}

\subsubsection{Participants}

 This study was conducted at
 Universidad Carlos III de Madrid, Spain, 
 with 45 participants enrolled in the Bachelor of Engineering program over three academic years (2020-2023). A control group consisting of 15 students from the 2017-2018 academic year, where no gamification elements were incorporated, was used for comparative analysis. During the academic years 2018-2019 and 2019-2020, the course was not taught due to low demand from students requesting enrollment in it.

\subsection{Course Implementation and Framework}

The innovative teaching methodology was deployed within the "Biomedical Applications of Nanotechnology" course, an advanced module for fourth-year students of the Bachelor's degree in Biomedical Engineering at 
UC3M, Spain.
This course, integral to the curriculum, is designed to provide students with 6.0 European Credit Transfer and Accumulation System (ECTS) credits, signifying its comprehensive nature and workload. 

\subsubsection{Curriculum Composition}

The course structure is meticulously crafted to blend foundational nanotechnology principles with their practical applications in the biomedical domain. Through a combination of lectures, team-oriented activities, and laboratory sessions, the course aims to build a strong understanding of nanotechnology's crucial role in solving biomedical challenges. A significant emphasis is placed on the conceptualization and development of nanotechnology-based devices and nanoparticles, particularly focusing on their clinical utility in diagnostic imaging and therapeutic interventions.

To ensure a well-rounded and effective learning experience, the course advocates for a prerequisite knowledge base in chemistry, materials science, biomedical instrumentation, and related fields. This foundational understanding is crucial for students to fully engage with and benefit from the advanced concepts and practices introduced in this course.  

\subsubsection{Education Objectives and Learning Outcomes}

The course aims to cultivate a deep conceptual and practical grasp of nanotechnology applications within biomedicine, fostering not only academic proficiency but also promoting key professional skills such as teamwork, critical thinking, and social responsibility. Through rigorous training, students are expected to attain a deep understanding of the potential of nanotechnology and its limitations in biomedical contexts, preparing them for future challenges and innovations in the field. 

\subsubsection{Pedagogical Strategies and Learning Modalities}

Adopting a dynamic and student-centered pedagogical approach, the course integrates traditional lectures with interactive seminars and practical laboratory sessions. This blend ensures a balanced delivery of theoretical knowledge and hands-on experience, encouraging active learning and critical engagement with the material. Prior to each session, students are tasked with preparatory assignments to maximize the effectiveness of the in-class activities. The course also incorporates a structural tutoring system to support the academic progress of the students and comprehensively address individual learning needs. 

\subsubsection{Assessment System}

The assessment framework of the "Biomedical Applications of Nanotechnology" course is designed to offer a multifaceted evaluation of student performance, synthesizing theoretical comprehension with practical application. The methodology is crafted to assess the adeptness of the students in applying nanotechnology principles to biomedical challenges, reflecting a dynamic synthesis of engaging activities and collaborative endeavors.

\begin{itemize}

    \item \textbf{Continuous Evaluation (CE) (60\% of Total Grade):}
    \begin{itemize}

        \item \textbf{Homework and interactive sessions (20\% of the CE grade):} Encompass assigned tasks and interactive sessions designed to reinforce theoretical concepts and foster practical application. Active participation and consistent attendance are crucial, as they significantly contribute to the engagement and assimilation of the course material. 

        \item \textbf{Laboratory experiences (30\% of the CE grade):} Practical sessions are integral, providing hands-on experience in nanotechnology applications within biomedicine. A minimum attendance of 80\% is requisite, with assessments comprising quizzes and concise reports to gauge proficiency in experimental procedures and data analysis. 
        
        \item \textbf{Midterm assessment \& Nanogames (20\% of the CE grade):} A pivotal assessment, the midterm examination, coupled with the Nanogames activity, evaluates the retention and the understanding of the course content by the students. The innovative Nanogames, detailed in Section~\ref{ss:nanogames}, encourage competitive and collaborative learning, enhancing conceptual understanding through gamified challenges. 

        \item \textbf{NanoTechStart project (30\% of the CE grade):} This collaborative venture, outlined in Section~\ref{ss:nanotechstart}, necessitates the integration of course concepts into a coherent proposal for a nanotechnology-based innovation. Emphasis is placed on effective communication and entrepreneurial acumen, aiming to simulate a real-world pitch to potential investors. 
        
    \end{itemize}

    \item \textbf{Final Exam (40\% of total grade):}
    
    The final examination encompasses a comprehensive assessment of the students' understanding of the course material, covering lectures, seminars, and laboratory sessions. Adherence to academic integrity is paramount, with stringent measures in place to ensure a fair and honest evaluation process. 
    
    \item \textbf{Extraordinary Exam:}
    
    An alternative examination pathway is available for students seeking to assess their overall course performance, offering a chance to re-evaluate their comprehension and application of the course material in its entirety. 
    
\end{itemize}

This assessment framework has been contrasted with a control group from the 2017-2018 academic year, which lacked the gamification components. This comparison aims to quantify the impact of gamification strategies on student engagement, motivation, and academic achievement.

\subsection{Midterm Exam \& Nanogames Engagement Activity}\label{ss:nanogames}

\begin{figure}[h]
	\centering
		\includegraphics[width=\textwidth]{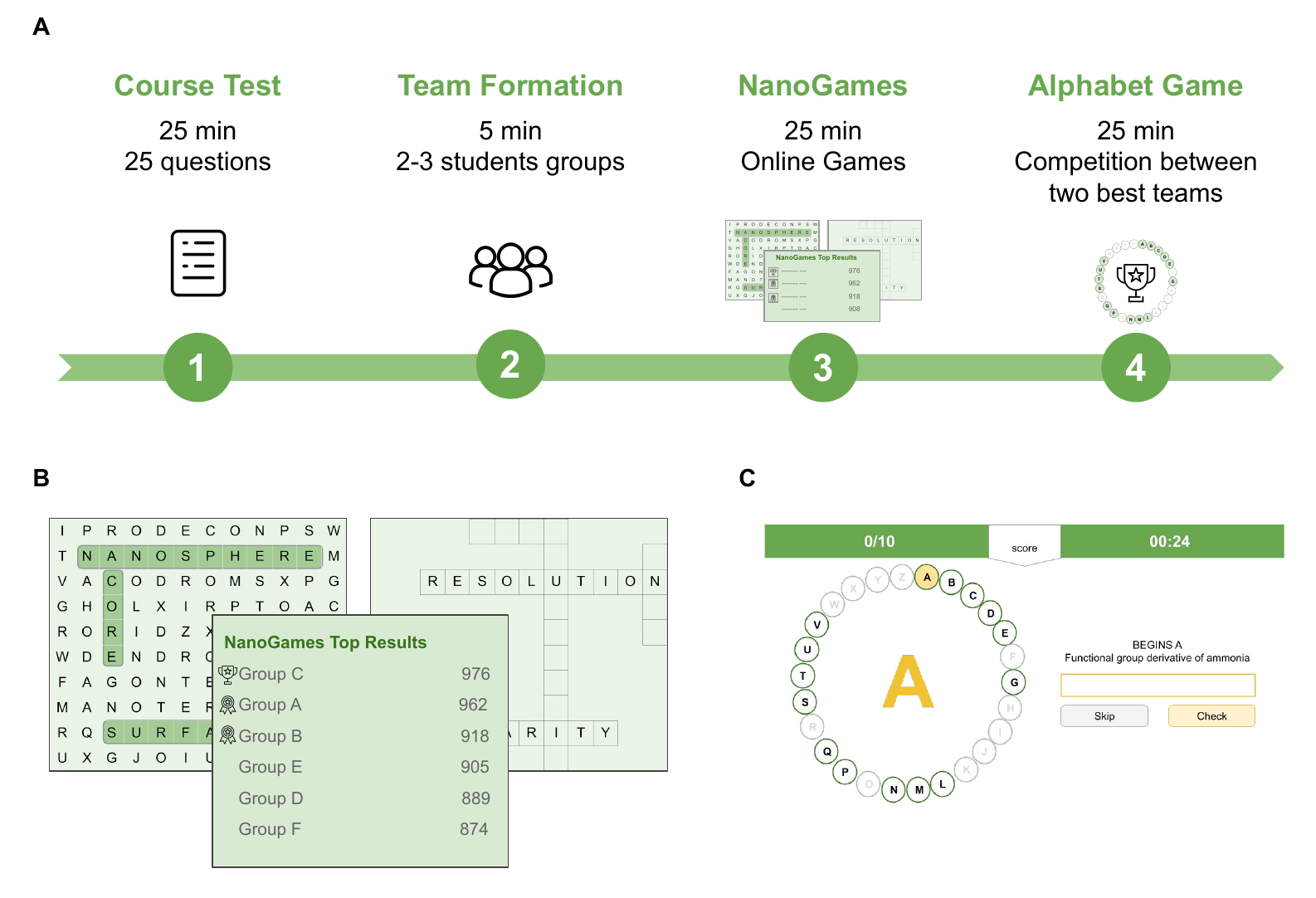}
	  \caption{A: Detailed illustration of the midterm evaluation structure, including the course test, team formation, Nanogames activities, and the final Alphabet game. Comprehensive information for each component is presented in the accompanying panel. B: Example of Nanogames activities, such as word searches and crosswords, along with the online leaderboard. C: Example panel for the Alphabet game.}\label{fig:nanogames}
\end{figure}

Incorporating the "Alphabet Game," this segment of the course assessment intertwines a traditional midterm examination with interactive, gamified elements designed to deepen the students' grasp of key concepts while providing a dynamic and engaging learning experience. This innovative approach leverages the principles of gamification to foster collaborative learning and enhance student motivation through competition and play. The structure and components of this activity are depicted in Fig.~\ref{fig:nanogames}A, illustrating the seamless integration of examination and game-based learning. 

\subsubsection{Examination Structure}

The midterm examination serves as the initial phase, assessing students on a comprehensive range of topics covered in the course, including theoretical foundations and practical applications explored during laboratory sessions. The examination format includes a variety of question types, such as true/false, multiple-choice, fill-in-the-blanks, and short definitions, aiming to evaluate students' understanding from multiple perspectives. Automated scoring facilitates immediate feedback, with the exam designed to be completed within a 45-minute timeframe. 

\subsubsection{Time Formation and Interactive Games}

Following the examination, students are stratified into teams based on their performance, with top scores assuming leadership roles. This team-based structure not only promotes peer learning but also encourages diverse interactions among students, breaking conventional group dynamics. The subsequent phase involves participation in a series of interactive, web-based games hosted on the Educaplay platform \cite{educaplay2023}. These activities, ranging from crosswords to matching games, are crafted to reinforce and apply course material in a stimulating format. Real-time leaderboards introduce a competitive edge (Fig.~\ref{fig:nanogames}B), enhancing engagement and motivation among students. 
Each activity is scored separately, and in between activities, students can view the current leaderboard displaying all group scores. This interactive segment is designed to last approximately 20 minutes. 

\subsubsection{The Alphabet Game: A Competitive Finale}

The top two teams from the Nanogames activity will advance to the final round, 'The Alphabet Game,' an exercise designed to simulate a competitive quiz format akin to popular television quiz shows (Fig.~\ref{fig:nanogames}C). This engaging and dynamic finale is structured to encompass a series of definitions relevant to the course's lexicon, ensuring a comprehensive review of key terms and concepts. The winning team will receive an additional +0.5 points over ten towards their midterm exam score, while the runners-up will be awarded an extra 0.25 points. The final round is allocated a total of 25 minutes. 

\subsubsection{Implementation and Feedback}

The entire Nanogames and Alphabet Game activities span approximately 100 minutes, offering a comprehensive and immersive educational experience. The orchestration of these activities lies with the educators, who not only conceptualize but also actively facilitate the event. This hands-on approach by educators is pivotal in assigning leadership roles among students, thereby nurturing essential skills such as teamwork, leadership, and active participation. 

Post-activity, a feedback mechanism is implemented through evaluation surveys. This critical step is instrumental in refining the educational approach for subsequent iterations and provides valuable insights into the participants' experiences. This feedback loop is crucial for continuous improvement and adaptation of the teaching methodology, ensuring that it remains effective, engaging, and in alignment with educational objectives.

\subsection{NanoTechStart}\label{ss:nanotechstart}

\begin{figure}[h]
	\centering
		\includegraphics[width=\textwidth]
            {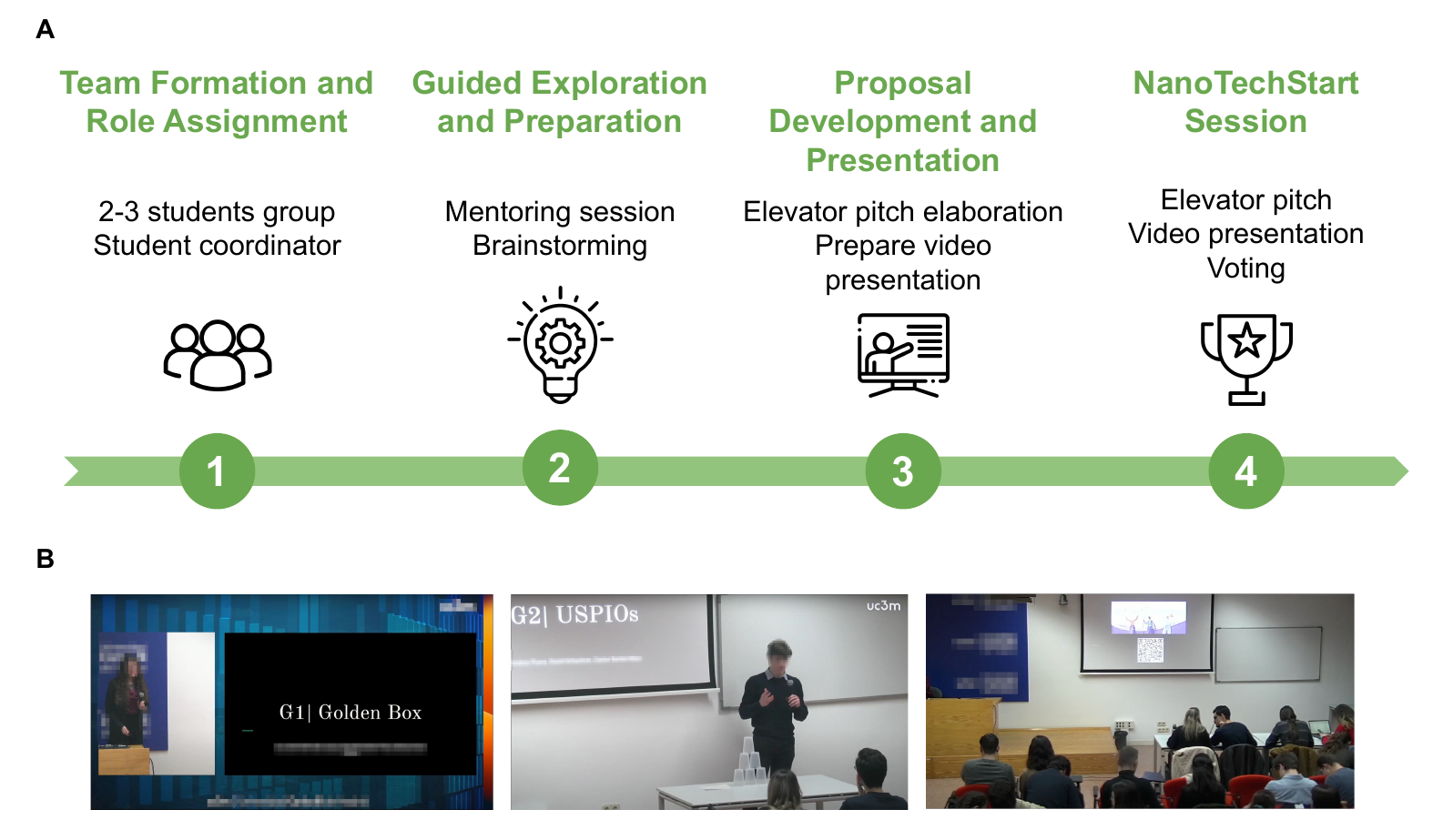}
	  \caption{A: NanoTechStart activity structure comprised of four parts, namely, the team formation and manuscript selection; presentation preparation; video presentation, and NanoTechStart session.
   B: NanoTechStart session. From left to right, welcome speech, elevator pitch and video presentation, and final vote and investor verdict.}\label{fig:nanotechstart}
\end{figure}

The NanoTechStart initiative, as part of our broader educational framework, aims to bridge theoretical knowledge with practical application, fostering an entrepreneurial mindset among students (Fig.~\ref{fig:nanotechstart}). Modeled after the HealthStart MadrI+D program, this activity encourages the creation of technology-driven startups within the health sector, originating from hospitals, research institutions, and universities, particularly within the Madrid community. 

\subsubsection{Team Formation and Role Assignment}

In the NanoTechStart activity, students are grouped into compact teams, typically comprising 2-3 members (Fig.~\ref{fig:nanotechstart}A), to foster close collaboration and effective communication. A unique aspect of this activity is the nomination of a student coordinator from within the group members, incentivized by a bonus of +0.5 points towards their overall grade for this role. This position is pivotal for steering the efforts of the team and ensuring cohesive progress throughout the project. Each team embarks on a journey of discovery by selecting cutting-edge technology from a curated database of nanotechnology articles provided by the instructors, aiming to explore its potential application within the biomedical field. 

\subsubsection{Guided Exploration and Preparation}

The initial phase of the NanoTechStart activity involves a guided exploration session, where teams receive personalized mentoring to delve deeper into their chosen technology. This session is instrumental in crystallizing the teams' understanding and perspective on the selected nanotechnology's application and its broader implications for healthcare. A brainstorming session follows, stimulating creativity and critical thinking as teams ponder the various dimensions of their project, setting the stage for the subsequent development of their proposals.

\subsubsection{Proposal Development and Presentation}

Following the exploratory session, teams are tasked with three core objectives: crafting a concise 'elevator pitch' that encapsulates the essence of their proposal, elaborating on the technological and product specifics, and preparing to advocate for their innovation to a panel of potential investors. The goal is to present a compelling narrative that not only demonstrates a deep understanding of the technology but also highlights its socio-economic benefits. The culmination of this effort is a video presentation, limited to 10 minutes, which serves as a comprehensive showcase of their proposal. These presentations are uploaded to a designated platform, with a submission deadline set one week before the final NanoTechStart session.

\subsubsection{The culminating Event: NanoTechStart Session}

The NanoTechStart session represents the pinnacle of this activity, simulating a real-world investment pitch environment. Teams, assuming the role of startups, present their innovations to a panel comprising faculty members and industry experts, effectively transforming the classroom into a venture capital pitch session. The session unfolds with a formal introduction of the teams and the jury, followed by the delivery of elevator pitches and video presentations. This interactive format not only enhances the learning experience but also allows for immediate feedback and engagement with the panelists. A democratic voting process ensues, determining the standout presentations, which are then duly recognized. The entire session is designed to be an immersive 100-minute experience with the potential to significantly impact the grades of the students based on their performance. The maximum possible score for this activity is 10.0.

\subsection{Data Collection Instruments}

This study adopts a multi-dimensional approach to evaluate the engagement and learning outcomes of participants, leveraging a blend of quantitative and qualitative data collection methods. The comprehensive analysis integrates feedback from educational activity surveys, performance evaluations, and end-of-course academic metrics to paint a holistic picture of the image of the pedagogical interventions. 

\subsubsection{Faculty Evaluation Surveys}

The faculty evaluation process is a critical component of our educational assessment strategy, designed to garner comprehensive feedback on teaching effectiveness and foster a constructive dialogue between students and educators (Table~\ref{tab:notas}). Conducted at the culmination of each semester, the evaluation employs a secure online platform accessible through the institutional intranet of the university, ensuring both confidentiality and ease of access for all participants. 

The survey is meticulously structured to evaluate various facets of teaching performance and pedagogical approach. Comprising eight meticulously crafted questions (Table~\ref{tab:encuestadocente}), the survey employs a five-point Likert scale, ranging from 'strongly disagree' (1) to 'strongly agree' (5). This format is designed to assess the instructional efficacy of the educators and solicit direct feedback from students regarding their learning experience. The dual focus not only aids in enhancing teaching methodologies but also strengthens the educational rapport between faculty and students, contributing to a more engaging and effective learning environment. 

Adhering to the highest standards of research ethics, the survey ensures complete anonymity for respondents, thereby encouraging candid and constructive feedback. The institutional commitment to ethical data handling and respondent anonymity underpins the integrity of the survey, fostering an environment where the students can freely express their views without fear of repercussion. 

To safeguard against multiple submissions and ensure the authenticity of the data, a unique MD5 hash is generated for each participant. This cryptographic measure precludes the possibility of duplicate responses while maintaining the anonymity of the participants. It is imperative to note that the MD5 hash, by its design, is a one-way transformation, ensuring that individual responses cannot be traced back to the respondents, thus upholding the principle of confidentiality. 

The aggregation of survey responses facilitates a comprehensive analysis while preserving individual anonymity. This aggregated data forms the basis for an in-depth understanding of teaching effectiveness, informing both internal review processes and strategic educational enhancements—the results, accessible to researchers post-evaluation period, provide invaluable insights for continuous improvement in teaching practices and curriculum development. 

\subsubsection{Student Engagement and Perception Surveys}

In order to assess the effectiveness and reception of the 'Nanogames' and 'NanoTechStart' initiatives, two distinct online surveys were deployed after each activity. These surveys, developed using Google Forms, were chosen for their accessibility and ease of dissemination among the student cohort, facilitating widespread participation. Conducted during the 2022-23 academic year, the surveys aimed to gauge students' satisfaction levels and perceived value of these gamified learning experiences in enhancing relevant skills.

The survey items (Table~\ref{tab:gameVStechstart}), designed to measure various aspects of student engagement and perception, were based on a five-point Likert scale ranging from 1 (strongly disagree) to 5 (strongly agree). This scale allowed for a detailed capture of students' attitudes towards the activities. To ensure the reliability and internal consistency of the survey instruments, a Cronbach's alpha coefficient analysis was conducted, resulting in a satisfactory alpha value of 0.925, indicating high reliability. 

Access to the surveys was restricted to students' institutional email accounts, a measure implemented to uphold the integrity of the data by preventing multiple submissions from a single participant. This approach also ensures the anonymity of respondents, encouraging candid feedback. Prior to distribution among the students, the survey itself was rigorously reviewed by two external experts to ensure its clarity, comprehensibility, and objectivity. The survey's design and implementation were carefully considered to align with best practices in educational research, ensuring the collection of meaningful and actionable data. This expert analysis prior to the survey release significantly bolstered the validity of our research approach.

\subsection{Statistical methods}

Statistical significance was determined using a p-value threshold of 0.05. Grades were analyzed using the Kruskal-Wallis test, while the Z-score calculator for two population proportions was employed to assess the difference in proportions. All values in the tables are expressed as $n$ (\%) or mean $\pm$ standard deviations. Cronbach's alpha coefficient was employed to measure the internal consistency of the questionnaires.


\section{Results}\label{s:results}

This section delineates the academic achievements of students throughout the 2020-2023 academic years in comparison to a control group from 2017-2018. The analysis encompasses student enrollment trends, performance metrics, and feedback from student surveys, thereby offering a holistic view of the educational impact over these periods.

\subsection{Student Performance}\label{ss:student_performance}

The analysis of student performance reveals a noteworthy increase in both enrollment and academic success rates in the Bachelor Engineering program. The period from 2020 to 2023 saw a significant rise in student numbers, with a total enrollment of 45 students, making a substantial growth compared to the 17 students in the 2017-2018 control group. This escalation in student numbers was accompanied by an improvement in academic outcomes, with a 100\% pass rate observed in the study group, contrasting with a 70.6\% pass rate in the control group. 

The data, as illustrated in Table~\ref{tab:notas}, encapsulates the enrollment figures and final grades across the academic years under consideration. A notable trend is the progressive improvement in the final grades, with the majority of students achieving higher scores in successive years. The year 2022-2023, in particular, stood out, with a significant proportion of students attaining grade A, underscoring the efficacy of the implemented educational strategies. 

The statistical analysis further substantiates the positive impact of these educational interventions, with a p-value of 0.00003 indicating a significant difference in the pass rates and academic achievements between the study and control groups. All students in the study group took the exam and 15 students (88.2\%) were in the control cohort (p-value=0.0735). These evidences point towards the successful enhancement of the student learning outcomes, attributable to the innovative pedagogical approaches adopted during this period. 

\begin{figure}
    \centering
    \includegraphics[width=\textwidth]{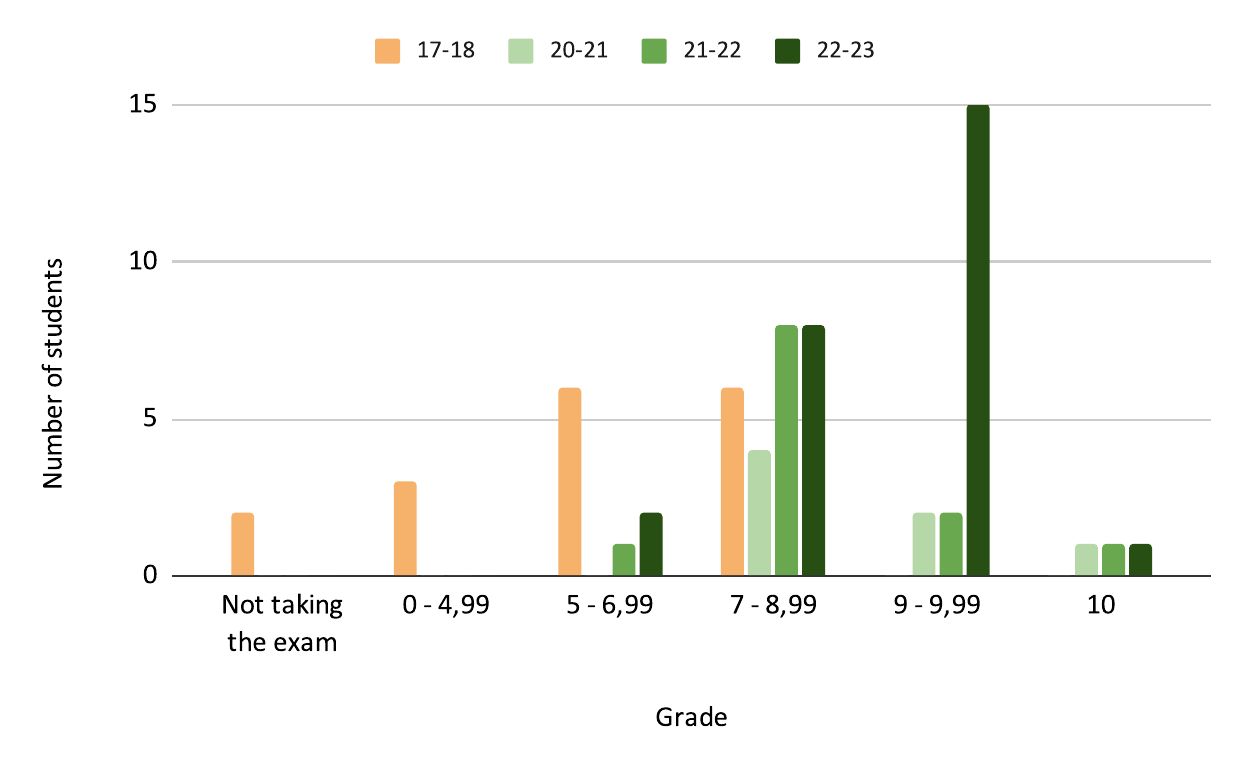}
    \caption{Comparative Analysis of Student Grades Over Four Academic Years (2017-2018 to 2022-2023). This graph illustrates the impact of the novel gamification-based pedagogical approach on student academic performance. The 2017-2018 academic year serves as a control period, during which traditional teaching methods were employed, while the innovative gamification methodology was implemented in the subsequent years. Notably, the introduction of gamification is associated with a marked improvement in student grades, demonstrating the efficacy of this approach in enhancing academic outcomes.}
    \label{fig:histogram}
\end{figure}

\begin{table}[h]
\caption{Student enrollment and final scores by academic year.}\label{tab:notas}
\begin{tabular*}{\textwidth}{@{\extracolsep\fill}lllllll}
\toprule%
\textbf{Academic Year} & & \textbf{17-18} & \textbf{20-21}     & \textbf{21-22}      & \textbf{22-23}      & \textbf{p-value}\footnotemark \\ 
\midrule
 \textbf{Number of students (N)}  &                & 17  & 7         & 12         & 26         & -       \\
\multicolumn{2}{@{}c@{}}{\hspace{-50pt} Students taking the exam} & 15 (88.2)  & 7 (100.0) & 12 (100.0) & 26 (100.0) &  0.0735       \\ 
\multicolumn{2}{@{}c@{}}{\hspace{-77pt} Students who pass} & 12 (70.6)  & 7 (100.0) & 12 (100.0) & 26 (100.0) & 0.0033       \\
\midrule
\textbf{Final Score } & \textbf{Final Score}  &  &  &            &            & \\
\textbf{(Numerical Scale)} & \textbf{(US Grade)}    &  &  &            &            & $<$0.0001 \\
\qquad 10.00        & A+ &  0 (0.0) & 1 (14.3)  & 1 (8.33)   & 1 (3.8)    &  \\
\qquad 9.00 - 9.99  & A  &  0 (0.0) & 2 (28.6)  & 2 (16.7)   & 15 (57.7)  &        \\
\qquad 7.00 - 8.99	& B  & 6 (35.3) & 4 (57.1)  & 8 (66.7)   & 8 (30.8)   &        \\
\qquad 5.00 - 6.99	& C  & 6 (35.3) & 0 (0.00)  & 1 (8.3)    & 2 (7.7)    &        \\
\qquad 0.00 - 4.99	& F  & 3 (17.6) & 0 (0.0)  & 0 (0.0)    & 0 (0.0)    &        \\
\qquad Not taking the exam	& -  & 2 (11.8) & 0 (0.0)  & 0 (0.0)   & 0 (0.0)  &        \\
\botrule
\end{tabular*}
\footnotetext[1]{p-value for the number of students compares the control group 17-18 and the last gamification year 22-23. P-value for the final scores was calculated for all the years.}
\end{table}

\subsection{Faculty survey}\label{ss:encuestas}

The subsection presents a comprehensive analysis of faculty survey data collected from students at the culmination of each semester spanning four academic years: 2017-2018, 2020-2021, 2021-2022, and 2022-2023 (Table~\ref{tab:encuestadocente}). The survey aimed to gauge students' perceptions of the course quality, a direct comparison with the overall degree program survey results that summarizes a total of 50 courses, and the effectiveness of the teaching methodology employed.

The survey data indicate a notable improvement in students' assessment of course quality over the years, with ratings escalating from 3.6 in 2017-2028 to a peak of 5.0 in the 2020-2021 academic year before slightly adjusting to 4.6 and 4.8 in subsequent years. This upward trajectory suggests a positive reception of the pedagogical enhancements implemented over this period. Conversely, the overall quality of the degree program remained relatively stable, with a consistent average ratio of 4.1, indicating a solid foundation in the broader educational offerings. 

An analysis of the teaching effectiveness (Table~\ref{tab:encuestadocente}), as perceived by students, revealed consistently high ratings across various dimensions, including the stimulation of learning, clarity of instruction, and the resolution of doubts, all exceeding the 4.0 threshold indicative of teaching excellence. These findings underscore the faculty's commitment to fostering an engaging and supportive learning environment. 

Feedback on student engagement and learning outcomes was overwhelmingly positive, with all assessed areas averaging above 4.0, except for the estimated weekly hours dedicated to the course. This exception may highlight a need to recalibrate course workload expectations. Notably, these results signify an enhancement in the learning experience within the Bachelor's program, as evidenced by comparative analysis with control group data. 

\begin{sidewaystable}
\scriptsize
\caption{Faculty survey filled by the students at the end of the semester. Results include the course and the entire degree program.}\label{tab:encuestadocente}
\begin{tabular*}{\textheight}{@{\extracolsep\fill}l|cc|cc|cc|cc}
\toprule%
\midrule
\textbf{Academic Year}                                                 & \multicolumn{2}{@{}c@{}|}{\textbf{17-18}} & \multicolumn{2}{@{}c@{}|}{\textbf{20-21}} & \multicolumn{2}{@{}c|}{\textbf{21-22}} & \multicolumn{2}{@{}c@{}}{\textbf{22-23}} \\
\midrule
\toprule
\textbf{Number of Students} & \multicolumn{2}{c|}{17} & \multicolumn{2}{c|}{7} & \multicolumn{2}{c|}{12} & \multicolumn{2}{c}{26} \\
\midrule
\textbf{Survey Participacion}                                                                              & \multicolumn{2}{c|}{ 7 (46.7)}  & \multicolumn{2}{c|}{7 (100.0)}       & 
\multicolumn{2}{c|}{7 (58.3)} & \multicolumn{2}{c}{12 (46.1)} \\
\quad - Overall average quality (course)                                                                   & \multicolumn{2}{c|}{3.6} & \multicolumn{2}{c|}{5.0} & \multicolumn{2}{c|}{4.6} & \multicolumn{2}{c}{4.8}\\
\quad - Overall average quality (entire program)                                                  & \multicolumn{2}{c|}{4.0} & \multicolumn{2}{c|}{4.1} & \multicolumn{2}{c|}{4.1} & \multicolumn{2}{c}{4.3} \\
\midrule
\textbf{Teacher Performance Questions} & \textbf{Course} & \textbf{Program} & \textbf{Course} & \textbf{Program} & \textbf{Course} & \textbf{Program} & \textbf{Course} & \textbf{Program} \\   
\quad \begin{tabular}{@{}l@{}}- The teacher stimulates learning appropriately                                                      \end{tabular} & 3.43 & 3.61 & 5.0 & 4.0 & 4.46 & 3.99 & 4.88 & 4.1           \\
\quad \begin{tabular}{@{}l@{}}- The teacher plans and coordinates the class\\ well and gives clear explanations\end{tabular} & 2.57 & 3.48 & 4.9 & 4.0 & 4.4 & 4.1 & 4.7 & 4.1 \\
\quad \begin{tabular}{@{}l@{}}- The teacher adequately solves doubts and\\ guides the students \end{tabular} & 3.14 & 3.63 & 4.9 & 4.1 & 4.6 & 4.2 & 4.8 & 4.2 \\
\quad \begin{tabular}{@{}l@{}}- Overall I am satisfied with the teacher's\\ teaching of the subject \end{tabular} & 3.00 & 3.58 & 5.0 & 4.0 & 4.6 & 4.1 & 4.7 & 4.3 \\
\midrule
\textbf{Feedback From Students}& & & & & & & \\
\quad \begin{tabular}{@{}l@{}}- Overall I am satisfied with the course \end{tabular}& 3.5$\pm$1.1 & 3.7$\pm$1.2 & 4.9$\pm$0.3 & 3.9$\pm$1.0 & 4.3$\pm$0.4 & 3.9$\pm$0.9 & 4.5$\pm$0.6  & 4.0$\pm$1.0     \\
\quad \begin{tabular}{@{}l@{}}- Evaluate the degree of coordination in the\\ course \end{tabular}& 3.2$\pm$0.7 & 3.7$\pm$1.1 & 4.5$\pm$0.5 & 3.9$\pm$1.0 & 4.3$\pm$0.4 & 4.0$\pm$0.9 & 4.5$\pm$0.8  & 3.9$\pm$1.0 \\
\quad \begin{tabular}{@{}l@{}}- Estimate the number of hours per week\\ dedicated to the course \end{tabular}& 3.2$\pm$1.2 & 2.8$\pm$0.9 & 3.1$\pm$0.6 & 2.9$\pm$1.0 & 2.1$\pm$0.6 & 2.6$\pm$0.9 & 2.4$\pm$0.7 & 2.6$\pm$0.9 \\
\quad \begin{tabular}{@{}l@{}}- Increase in knowledge, competencies and/or\\ skills acquired \end{tabular}& 3.7$\pm$1.0 & 3.7$\pm$1.1 & 4.6$\pm$0.7 & 3.9$\pm$0.9 & 4.1$\pm$0.3 & 3.9$\pm$0.9 & 4.2$\pm$0.9 & 3.9$\pm$0.9 \\ 
\botrule
\end{tabular*}
\end{sidewaystable}

\subsection{Student Gamification Experience Surveys}\label{ss:encuestas_actividades}

Table~\ref{tab:gameVStechstart} provides feedback and satisfaction results for the newly proposed activities. The aggregated data from the surveys indicate a high level of overall satisfaction among participants, with 93.7\% of Nanogames and 76.2\% of NanoTechStart participants expressing overall satisfaction with their respective activities (p-value=0.1499). Notably, a significant proportion of students acknowledged an increase in subject knowledge and soft skills, attributing it to their engagement in these gamified learning environments.

Specifically, 81.2\% of Nanogames and 71.4\% of NanoTechStart participants reported a perceived increase in their understanding of the subject matter (p-value=0.4902). All Nanogames participants (100\%) and a majority of NanoTechStart participants (81.0\%) felt an improvement in their soft skills (p-value=0.0643), underscoring the role of gamification in fostering essential non-technical competencies. Both activities were praised for their optimal duration and effective facilitation of communication between educators and students, with unanimous approval from Nanogrames participants and above 95\% from NanoTechStart participants (p-value=0.3789). 

\begin{table}[h]
\scriptsize
\caption{Survey of students rating the two gamification activities Nanogames and NanoTechStart for the academic year 2022-2023.}\label{tab:gameVStechstart}
\begin{tabular*}{\textwidth}{@{\extracolsep\fill}l|cc|cc}
\toprule%
\textbf{Gamification activity} & \multicolumn{2}{@{}c|@{}}{Nanogames} & \multicolumn{2}{@{}c@{}}{NanoTechStart} \\
\midrule
\textbf{Student activity satisfaction} & \textbf{Score} & \textbf{Score ($>4.0$)} & \textbf{Score} & \textbf{Score ($>4.0$) }\\
\quad - Successfully increased knowledge\\ \quad of the subject & 4.2$\pm$0.9 & 13/16 (81.2\%) & 4.0$\pm$0.9 & 15/21 (71.4\%) \\
\quad - Enhanced "soft skills" & 4.4$\pm$0.6 & 16/16 (100.0\%) & 4.2$\pm$0.9 & 17/21 (81.0\%) \\
\quad - Appropriate duration of the activity & 4.8$\pm$0.4 & 16/16 (100.0\%) & 4.5$\pm$0.6 & 21/21 (100.0\%) \\
\quad - Effective communication between \\ \quad teacher and students & 4.8$\pm$.04 & 16/16 (100.0\%) & 4.5$\pm$0.6 & 20/21 (95.2\%) \\
\quad - Overall satisfaction with the activity & 4.6$\pm$0.6 & 15/16 (93.7\%) & 4.1$\pm$1.0 & 16/21 (76.2\%) \\
\botrule
\end{tabular*}
\end{table}

\subsection{Gamification Theory Results}

\begin{figure}[h]
	\centering
		\includegraphics[width=\textwidth]{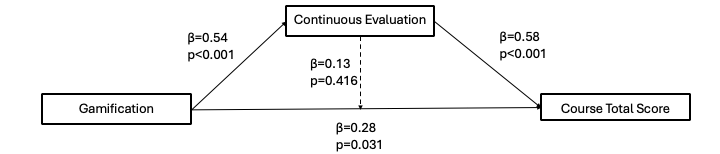}
	  \caption{Results of the moderated-mediation analysis. The $\beta$ coefficients and their statistical significance p-values are displayed for the direct, mediation, and moderation paths, respectively.}\label{fig:MMmodelresults}
\end{figure}

Finally, we include the results of the mediation-moderation process that we have analyzed according to the "Theory of Gamified Learning"~\cite{D_Landers_2014}. This approach allows us to systematically evaluate the influence of gamified elements on learning outcomes through their direct and indirect effects on learner behaviors and attitudes. By adhering to this theoretical framework, we aim to provide a clear and structured analysis of how gamification impacts educational effectiveness. We have focused the analysis on the relationships depicted in the mediation-moderation model, with continuous evaluation as the mediator and/or moderator, see Fig.~\ref{fig:MMmodelresults}. The correlation of these variables is presented in Table~\ref{tab:MMcorrelation}, and the moderated-mediation results after analyzing the model in Fig.~\ref{fig:MMmodelresults} are provided in Table~\ref{tab:MMregression} and Table~\ref{tab:MMeffect}.

\textbf{Hypothesis 1} stated that the positive effect of gamification on the course total score will be mediated by the continuous evaluation performance. The regression analysis in Table~\ref{tab:MMregression} shows a significant positive effect of gamification on the mediating continuous evaluation variable, with a coefficient $\beta$ of 0.54 ($p < 0.001$1, see Table~\ref{tab:MMregression}), which is also reflected in the significant effect ($\beta=0.58$, $p<0.001$) of the mediating variable on the dependent one, the course total score. In this case, the data supports Hypothesis 1 as the mediation effect of the continuous evaluation is significant, compared to only the direct effect of gamification over the course total score ($\beta=0.28$, $p=0.031$).

\textbf{Hypothesis 2} proposed a strong relationship between the performance of the students in the continuous evaluation and the final course score. From the data in Table~\ref{tab:MMregression}, continuous evaluation positively affects the course total score, with a coefficient $\beta=0.58$ ($p < 0.001$). This result suggests a strong relationship between continuous evaluation performance and the final course score, which supports Hypothesis 2, indicating that higher continuous evaluation scores are associated with higher overall course scores.

\textbf{Hypothesis 3} implied that the continuous evaluation performance will moderate the relationship between gamification and total course score. The interaction effect of gamification and continuous evaluation on the course total score is examined. The coefficient for the interaction term ($\beta=0.13$, $p = 0.416$) is not statistically significant. Therefore, Hypothesis 3 is not supported by the data, as continuous evaluation does not significantly moderate the relationship between gamification and course total score.

The overall model in Fig.~\ref{fig:MMmodelresults} explains a significant proportion of the variance in continuous evaluation ($R^2 = 0.29$, $p<0.001$) and in the course total score ($R^2 = 0.58$, $p < 0.001$). These results highlight the substantial direct impact of gamification on both continuous evaluation and course total score, while the indirect and moderation effects of continuous evaluation were not significant.

\begin{table}[h]
\scriptsize
\caption{Descriptive and correlation analysis of the variables in the moderation and mediation process.}\label{tab:MMcorrelation}
\begin{tabular*}{\textwidth}{@{\extracolsep\fill}l|cccccc|cc}
\toprule
 & N & Mean & Median & SD & Min. & Max. & CE (\%) & CT (\%)\\
\botrule
Gamification & 41 & - & - & - & - & - &  &  \\ 
CE (\%) & 41 & 0.89 & 0.91 & 0.08 & 0.66 & 0.98 & 0.01 &  \\
CT (\%) & 41 & 0.8 & 0.84 & 0.15 & 0.4 & 0.97 & 0.01* & 0.02 \\
\botrule
\end{tabular*}
\footnotesize{Notation: Gamification, normal evaluation=0, gamified evaluation=1. Continuous evaluation (CE) and Course Total (CT) are all percentages. *$p<0.001$.}
\end{table}

\begin{table}[h]
\scriptsize
\caption{Regression results for the tests regarding the moderated-mediation model with CE as mediator and moderator (see Fig. \ref{fig:MMmodelresults}).}\label{tab:MMregression}
\begin{tabular*}{\textwidth}{@{\extracolsep\fill}lccc}
\toprule
Model & $R^2$ &Coefficient & Standard error \\
\botrule
Effect on CE	& 0.29** & - & - \\
\quad Gamification $\rightarrow$ CE  & & 0.54**	& 0.02\\
Complete Model	     & 0.58**   & - & - \\
\quad CE $\rightarrow$ CT            & & 0.58**	& 0.25 \\
\quad Gamification $\rightarrow$ CT	 & & 0.28*	& 0.04\\
\quad Gamification x CE $\rightarrow$ CT & & 0.13	& 0.5\\
\botrule
\end{tabular*}
\footnotesize{Notation: Gamification, normal evaluation=0, gamified evaluation=1. Continous evaluation (CE) and Course Total (CT) are all percentages. *$p<0.05$, **$p<0.001$.}
\end{table}

\begin{table}[h]
\scriptsize
\caption{Conditional direct and indirect effects of the gamification condition on the course total scores.}\label{tab:MMeffect}
\begin{tabular*}{\textwidth}{@{\extracolsep\fill}llcc}
\toprule
Model &  & Effect & Standard error \\
\botrule
Direct Effects & Gamification $\rightarrow$ CT & 0.09* & 0.04 \\
Indirect Effect	& Gamification $\rightarrow$ CE $\rightarrow$ CT & 0.10* & 0.03 \\
Total Effect & Direct + Indirect & 0.19** & 0.04 \\
\botrule
\end{tabular*}
\footnotesize{Notation: Gamification, normal evaluation=0, gamified evaluation=1. Continous evaluation (CE) and Course Total (CT) are all percentages. *$p<0.05$, **$p<0.001$.}
\end{table}

\section{Discussion}

This study aimed to investigate and quantify the beneficial impacts of gamification on student engagement and motivation, as introduced in Section~\ref{ss:objectives}. Concurrently, the gamified experience offered by the proposed Nanogames and NanoTechStart activities was intended to assess students' collaborative and interactive abilities, while also enriching their understanding of real-world scenarios in the context of developing their own business projects. Thus, the discussion section is structured to align with these initial objectives.

\subsection{Impact of Gamification on Academic Outcomes in Engineering Education}

The incorporation of gamified elements into the engineering curriculum has markedly enhanced academic outcomes, as evidenced by increased student engagement and improved performance metrics. This aligns with existing literature that underscores the benefits of gamification in educational settings \cite{D_Lo2020, D_Zainuddin2020, D_Jurgelaitis2019, D_Jurgelaitis2018}. For instance, studies have highlighted the efficacy of gamified e-quizzes in enhancing student learning and engagement, noting significant improvements in performance scores across various groups \cite{D_Jurgelaitis2018, D_Jurgelaitis2019}. This phenomenon underscores the potential of gamification to transform traditional educational paradigms, making learning more interactive and engaging, thereby fostering deeper comprehension and retention of complex engineering principles~\cite{D_Lo2020}.

Our observations align with the broader educational research that suggests challenges and competition within a gamified context can significantly enhance learning outcomes \cite{D_Hamari2016}. This is attributed to the increased engagement and active participation it fosters among students. By making learning more enjoyable and collaborative, gamification encourages a more profound and meaningful educational experience. This not only augments academic performance but also facilitates the development of essential soft skills, such as teamwork and communication, which are critical in the professional engineering landscape \cite{D_Larson2020}.

Nevertheless, recognizing the diversity in student responses to gamification is crucial. The effectiveness of such educational strategies can vary significantly across different learning profiles, necessitating adaptive and personalized approaches to maximize their impact. Moreover, the integration of gamification elements should be approached with caution, as their influence on learning outcomes can be multifaceted and not universally positive \cite{D_Toda2018, D_Koivisto2019}.

Despite our research indicating a boost in overall student performance, ongoing exploration, and refinement of gamification approaches will further clarify their contribution to improving educational results and shaping the future of engineering education. 

\subsection{Fostering Leadership and Active Engagement in Engineering Education}

The strategic integration of gamification within engineering pedagogy extends beyond mere academic enrichment, serving as a catalyst for cultivating leadership qualities and active participation among students \cite{D_Larson2020}. This paradigm shift towards a more interactive and participatory learning environment is evidenced by the transformative impact of gamification strategies such as Nanogames and NanoTechStart, which necessitate and nurture student initiative, decision-making, and collaborative problem-solving. This engagement in the learning process not only enhances students' technical awareness but also impregnates them with a sense of ownership and accountability towards their educational journey, thereby spreading the seeds of leadership. 

In alignment with the Econplus Champions League model discussed by Murillo et al. (2021) \cite{D_Murillo2021}, where competition among student teams is leveraged to create a dynamic and gameful learning experience, our approach similarly empowers students through active involvement in the design and execution of gamified activities. This metho\-do\-logical framework not only reinforces students' engagement but also instills leadership competencies as they navigate through the gamified challenges and opportunities, thus mirroring real-world professional scenarios. 

The positive interplay between gamified learning strategies and enhanced student performance, as corroborated by existing literature \cite{D_Hamari2016, D_Zainuddin2020}, underscores the efficacy of integrating game-based elements in fostering an engaging, competitive, and collaborative learning environment. This approach not only renders the learning process more enjoyable but also catalyzes a shift from passive reception to active contribution and leadership within the educational setting. 

Moreover, our findings resonate with the broader pedagogical discourse, suggesting that gamification can significantly elevate the educational experience by enriching subject knowledge and polishing soft skills. This is evident from the positive reception of the Nanogames and NanotechStart activities by the students, as demonstrated by the survey results presented in this study (Table~\ref{tab:gameVStechstart}). Encouraging students to assume leadership roles and actively contribute to course activities transforms them from mere knowledge consumers to co-creators of their learning experience. 

This pedagogical strategy aligns with the emerging educational paradigms that advocate for a more student-centered approach \cite{D_Hao2023}, where learners are not only recipients of knowledge but also active participants in their educational development. By fostering an environment that encourages leadership, collaboration, and active involvement, we prepare students not just for academic success but for the multifaceted challenges of the professional world. 

As we continue to refine and expand upon these gamification strategies in future iterations of the course, it is imperative to explore further their potential in nurturing leadership qualities and enhancing student engagement in engineering education. This ongoing exploration will contribute to the evolving landscape of STEM education, ensuring that it remains responsive to the needs and aspirations of future engineers. 

\subsubsection{The Effect of Gamification Tools as Motivators}

The integration of gamification tools within the engineering curriculum has significantly demonstrated their role as powerful motivators, enhancing student engagement and fostering a passion for learning. By embedding elements synonymous with gaming, such as competition, achievement, and rewards, these tools have transformed conventional learning paradigms into dynamic and captivating experiences. Research, such as the study by Zainuddin et al. \cite{D_Zainuddin2020}, has shown that the introduction of game-like features like points, badges, and leaderboards in educational contexts significantly boosts student engagement through gamification. The strategic use of Nanogames and NanoTechStart activities leverages the motivational allure of gaming elements, thereby amplifying students' perseverance, effort, and enjoyment in their academic pursuits \cite{D_Jurgelaitis2018,D_CozarGutierrez2016}. This transition towards an interactive and enjoyable learning atmosphere not only encourages deeper engagement with the material but also fosters improved comprehension and retention of course content, which is pivotal for the success of the educational process \cite{D_Shi2014, D_CozarGutierrez2016}. 

However, it is crucial to acknowledge the varied motivational triggers among students, emphasizing the need for inclusive and adaptable gamification strategies that cater to diverse preferences and learning styles. Echoing the insights from \cite{D_Hallifax2020}, customizing gamification to align with both player types and motivational profiles can significantly enhance intrinsic motivation and reduce demotivation, thereby surpassing the effectiveness of one-size-fits-all approaches. As we continue to refine and expand the repertoire of gamification techniques, their sustained effectiveness as motivators will ensure they remain relevant to the evolving educational requirements and interests of engineering students. 

Additionally, the analysis grounded in the "Theory of Gamified Learning," reveals that gamification has a significant direct impact on both continuous evaluation and overall course scores. 
Similarly to what \cite{D_Landers_2014} and \cite{D_Landers_2014}, we motivated our students with leaderboards and competitive activities, which proved to have a positive impact on the final course score.
However, we found that continuous evaluation does not mediate or moderate the relationship between gamification and course total scores in a statistically significant manner. Previous results from \cite{D_MolinerHeredia_2023} indicated that implementing gamification in continuous assessment positively impacted attendance and engagement but had little to negligible effect on student performance improvement. These findings underscore the importance of gamification as a direct enhancer of educational outcomes, while the role of continuous evaluation as a mediator or moderator requires further investigation.

\subsubsection{Foster Social Needs and Economic Development Awareness}

The integration of gamification into engineering education transcends mere academic performance enhancement, emerging as a potent tool to cultivate awareness of social needs and economic development. This paradigm is supported by the notion that academics and industry should collaboratively forge workplace-relevant programs, as suggested by \cite{D_Pang2019}. Through initiatives like NanoTechStart, students are encouraged to consider the broader implications of engineering solutions, particularly their societal and economic impacts. This educational model is reinforced by a comprehensive, university-wide entrepreneurship framework aimed at invigorating students' entrepreneurial spirit, mirroring the observations by Mei et al.~\cite{D_Mei2022}. 

Such an educational approach not only aligns with but also amplifies the call for engineers to be socially responsible and ethically mindful. By immersing students in gamified simulations of real-world challenges, they are encouraged to critically evaluate the societal and economic considerations that underpin engineering decisions. This immersive pedagogy fosters a profound comprehension of engineering's role in addressing global challenges, priming students to conceive solutions that are technically sound, socially beneficial, and economically sustainable. 

As the engineering discipline continues to evolve, the integration of social responsibility and economic awareness into the curriculum becomes increasingly pivotal. By continuing to harness gamification as a conduit for engaging students with these vital issues, we can nurture a new cadre of engineers who are not only technically proficient but also socially conscious and globally aware.



\subsection{Limitations and Future Recommendations}

While the integration of gamification into engineering education has yielded promising outcomes, highlighting increased student engagement and improved academic performance, it is imperative to recognize the limitations inherent to this study and offer recommendations for future research and educational practices. One significant limitation is the contextual specificity of the research, conducted within a single engineering discipline at a particular institution, potentially limiting the universality and applicability of the findings across different educational contexts. To address this, future research should aim to replicate and extend this study across varied academic settings and disciplines, thereby validating the generalizability and adaptability of the gamification framework proposed.

Additionally, the study acknowledges the diverse preferences and reactions of students toward gamification strategies, underscoring the importance of developing a multifaceted approach that accommodates a spectrum of learning styles and preferences. This necessitates the incorporation of personalized and flexible gamification designs to cater to the varied needs of students, enhancing the inclusivity and effectiveness of gamified learning environments.

The study also points to the limitation posed by the brevity of the faculty survey, which might have constrained a more comprehensive understanding of student engagement and motivations. Future research could benefit from employing a more extensive array of measurement and data-collection tools to deepen the insights into student perceptions and engagement, thereby enriching the findings and implications of gamification in education.

Moreover, the investigation into the short-term impacts of gamification reveals a gap in understanding the long-term effects of such educational strategies on students' academic trajectories and lifelong learning habits. Longitudinal studies are thus recommended to explore the enduring influences of gamified learning on students' educational and professional development. 

Lastly, the study emphasizes the dynamic nature of technology and educational theories, advocating for continuous updates and innovations in gamification strategies and educational technologies. Educators and researchers are encouraged to stay abreast of the latest developments in the field, ensuring that gamified learning experiences remain relevant and effective in meeting the evolving educational needs and preferences of future generations of engineers. 

\section{Conclusions}

This study shows how gamification can transform engineering education, making it more immersive, interactive, and effective. The empirical evidence gathered from our investigation corroborates the assertion that meticulously curated gamification strategies, when seamlessly integrated into the curriculum, substantially amplify student engagement, motivation, and scholarly success. The additional benefits observed in student leadership, teamwork, and professional competencies further affirm that the ramifications of gamification transcend mere academic performance, fostering a comprehensive developmental trajectory for aspiring engineers.

Despite the promising results, the implementation of gamification in engineering education requires careful consideration of various factors, including student diversity, the availability of technological infrastructure, and overarching educational objectives. As the educational landscape continues to evolve, gamification emerges as a pivotal methodology to address the needs of forthcoming generations of engineers. 

This exploration into the efficacy of gamification within engineering education underscores the necessity for continued research and practical application to refine these pedagogical strategies. It is imperative to investigate further and harness the potential of gamification to enrich engineering education globally, ensuring it remains adaptive, student-centered, and responsive to the ever-changing digital landscape.

\section*{Ethical Approval}
Not applicable

\section*{Informed Consent}
Not applicable

\section*{Statement Regarding Research Involving Human Participants and/or Animals}
Not applicable

\section*{Consent to Participate}
Not applicable

\section*{Consent to Publish}
Not applicable

\section*{Funding}
Not applicable

\section*{Author's Contribution}
Not applicable

GRM: 
Conceptualization, Methodology, Writing – original draft, Writing – review \& editing, Data curation. 
CF-B: 
Writing – review \& editing.
AM-B: 
Conceptualization, Methodology, Writing – original draft, Writing – review \& editing, Project administration.

\section*{Competing Interests}
The authors declare that they have no competing interests.

\section*{Availability of data and materials}
The datasets used and/or analyzed during the current study are available from the corresponding author upon reasonable request.






\bibliography{references_journal.bib}

\end{document}